\documentclass{PoS}
\usepackage{amsmath, amssymb}

\title{The magnetic susceptibility in QCD}

\ShortTitle{The magnetic susceptibility in QCD}

\author{\speaker{Claudio Bonati}\\
        Dipartimento di Fisica dell'Universit\`a di Pisa and 
        INFN - Sezione di Pisa, Largo Pontecorvo 3, I-56127 Pisa, Italy\\
        E-mail: \email{bonati@df.unipi.it}}

\author{Massimo D'Elia\\
        Dipartimento di Fisica dell'Universit\`a di Pisa and 
        INFN - Sezione di Pisa, Largo Pontecorvo 3, I-56127 Pisa, Italy\\
        E-mail: \email{delia@df.unipi.it}}

\author{Marco Mariti\\
        Dipartimento di Fisica dell'Universit\`a di Pisa and 
        INFN - Sezione di Pisa, Largo Pontecorvo 3, I-56127 Pisa, Italy\\
        E-mail: \email{mariti@df.unipi.it}}

\author{Francesco Negro\\
        Dipartimento di Fisica dell'Universit\`a di Genova and INFN - Sezione di Genova,
        Via Dodecaneso 33, I-16146 Genova, Italy\\
        E-mail: \email{fnegro@ge.infn.it}}

\author{Francesco Sanfilippo\thanks{Now at: School of Physics and Astronomy, University of Southampton,
        Southampton SO17 1BJ, United Kingdom.}\\
        Laboratoire de Physique Th\'eorique (Bat. 210) Universit\'e Paris SUD, 
        F-91405 Orsay-Cedex, France\\
        E-mail: \email{francesco.sanfilippo@th.u-psud.fr}}

\abstract{
Recently much work has been devoted to the study of QCD coupled to a 
background magnetic field. Strongly interacting matter acts as a magnetic medium
and it is natural to study the properties of this medium, in particular to 
understand if it behaves like a diamagnetic or a paramagnetic material.
A serious difficulty in studying these properties by means of LQCD simulations
is the quantization of the magnetic field in a toroidal geometry. We will expose 
a method to overcome this difficulty and we will present data obtained for the 
$N_f=2$ theory that show that the QCD medium is paramagnetic.
}

\FullConference{31st International Symposium on Lattice Field Theory - LATTICE 2013\\
		July 29 - August 3, 2013\\
		Mainz, Germany}

\begin{document}

\section{Introduction}

Magnetic field of intensities ranging from $10^{10}$ to $10^{15-16}\,\mathrm{T}$ are 
expected to be present in such disparate environments as compact astrophysical objects 
(for magnetars see \emph{e.g} \cite{magnetars}), the early Universe (see \emph{e.g} 
\cite{cosmo}) and heavy ion collisions (see \emph{e.g.} \cite{heavy}). Since a 
magnetic field of intensity $10^{15-16}\,\mathrm{T}$ corresponds to 
$|e|B\sim 1\,\mathrm{GeV}^2$, it is clear that its interaction 
can significantly affect the properties of strongly interacting matter. 
As a consequence, in the last few years the interplay between strong and 
electromagnetic interactions have received much attention (for a comprehensive 
review see the volume \cite{lecnotmag}).

For ordinary materials, the computation of the reaction to an external magnetic 
field is a standard problem of condensed matter physics. For non ferromagnetic 
media and small magnetic fields, the induced polarization is linear in the intensity
of the external field and a quantitative measure of the reaction of the material is 
the magnetic susceptibility. It appears natural to ask the same type of fundamental questions 
for the strongly interacting matter: does it react linearly to external magnetic fields? 
If this is the case, what is the value of its magnetic susceptibility? In particular, 
is it a paramagnetic or a diamagnetic medium? Despite the simplicity and clear-cut 
nature of these questions, it is nontrivial to answer them. 

The standard tool for studying non-perturbative aspects of QCD dynamics 
is the lattice formulation of the theory and it is not difficult to add an external 
magnetic field to the discretized theory. However, in a toroidal geometry (the one 
usually adopted in simulations to reduce finite size effects), the magnetic field values 
are not arbitrary but get quantized.
Intuitively this is related to the fact that when applying the Stokes theorem on 
a compact manifold without boundary the result must be independent of the surface 
used for the flux computation. To enforce this independence we have to impose 
a relation between the admissible magnetic fluxes and the smallest electrical charge 
present in the theory. 
In the QCD case the smallest charge is $q=|e|/3$ and, assuming 
$\mathbf{B}=B\,\mathbf{\hat{z}}$ and a $3D$ toroidal manifold, one gets the 
quantization condition \cite{b}
\begin{equation}\label{bquant}
|e| B = {6 \pi b}/{(\ell_x \ell_y)}\ ,  
\end{equation}
where $\ell_x, \ell_y$ are the periods of the torus in the $x, y$ directions and 
$b\in\mathbb{Z}$. This quantization condition is the main obstruction to a 
simple lattice answer to the previous questions.

In the following we will present the method developed in \cite{cffmm} to overcome 
these difficulties and the results obtained by applying it to the case of $N_f=2$ 
staggered fermions: strongly interacting matter at finite temperature behaves 
as a linear paramagnetic medium and, near deconfinement, its magnetic susceptibility 
is of the same order of magnitude of that characterizing typical strongly 
paramagnetic ordinary materials (like \emph{e.g} liquid oxygen).

\section{The method}

The magnetic susceptibility and, more generally, all the magnetic properties of a 
(homogeneous) medium, are related to the change of the free energy density 
$f=F/V$ in presence of an external magnetic field:
\begin{equation}\label{freediff}
\Delta f(B,T) = - \frac{T}{V} \log \left( \frac{Z(B,T,V)}{Z(0,T,V)} \right)\ ,
\end{equation}
where $Z = \exp(-F/T)$ is the partition function. Since free energies are 
notoriously difficult to compute by means of numerical simulations, 
the standard procedure to evaluate magnetic susceptibilities in condensed 
matter simulations is to study the expectation value of the second derivative
of $f$, which is a much better behaved observable than the magnetic free energy 
density. This is however not possible in the present setting, since the 
magnetic field is quantized and, as a consequence, derivatives with respect 
to $B$ of Eq.~(\ref{freediff}) are not well defined.

The basic idea of the method introduced in \cite{cffmm} is to extract the magnetic 
susceptibility and the other magnetic properties directly from the behaviour of 
free energy differences $f(b_2)-f(b_1)\equiv f(B_2)-f(B_1)$ (where $b_2$ and 
$b_1$ are integers), which are computed by using the elementary formula 
\begin{equation}\label{intfb}
f(b_2) - f(b_1) = \int_{b_1}^{b_2} \frac{\partial f(b)}{\partial b} \mathrm{d} b\ ,
\end{equation}
with the integrand function being evaluated on a grid of points in the interval 
$[b_1, b_2]$ (grid that have to be fine enough for the errors associated to the 
numerical integration to be under control). In order to follow this strategy 
we have to analytically continue the function $f(b)$, which is properly 
defined only for $b\in\mathbb{Z}$, on the whole real axis, which is done 
in the following way.

An external magnetic field is introduced in Lattice QCD simulations by 
adding to the $SU(3)$ links variables $U_{\mu}(n)$ the non dynamical $U(1)$ 
phases $u_{\mu}(n)$ associated to the magnetic field, \emph{i.e.} with the 
replacement $U_{\mu}(n)\to u_{\mu}(n)U_{\mu}(n)$. A simple choice for  
the $U(1)$ phases corresponding to $\mathbf{B}=B\, \mathbf{\hat{z}}$ is \emph{e.g.}
\begin{equation} \label{u1field}
\begin{aligned}
& u_y^{(q)}(n) = e^{i\, a^2 q B\, n_x} & 
&\Big(=\ e^{i\, 2 \pi b\,  n_x / (L_x L_y)}\ \mathrm{for\ the\ } u\ \mathrm{flavor}\Big)\\
& u_x^{(q)}(n)|_{n_x = L_x} = e^{-i\, a^2 q L_x B\, n_y} & 
&\Big(=\ e^{-i\, 2 \pi b\, n_y / L_y}\ \mathrm{for\ the\ } u\ \mathrm{flavor} \Big)
\end{aligned} 
\end{equation}
and $u_{\mu}^{(q)}(n)\equiv 1$ otherwise. In this expression $q$ is the charge of 
the considered flavour, $L_x, L_y$ are the lattice extents in the $x,y$ directions, 
$a$ is the lattice spacing and $1\le n_{\mu}\le L_{\mu}$. The magnetic field in the 
$\hat{z}$ direction associated to the phases Eq.~(\ref{u1field}) is uniform only if the 
quantization condition Eq.~(\ref{bquant}) is respected, otherwise a singularity 
analogous to a Dirac sting is present in the continuum limit. The analytical 
continuation required for the application of Eq.~(\ref{intfb}) is obtained by 
removing the requirement that $b\in\mathbb{Z}$ in the expressions in 
Eq.~(\ref{u1field}). Since we work on finite lattices the free energy density $f(b)$ 
is then an analytic function of $b$ and Eq.~(\ref{intfb}) can be safely applied.

We emphasize that $\partial f(b)/\partial b$ defined in such a way is not related 
in any direct way to the magnetization of the system: the value of 
$\partial f(b)/\partial b$ (also for integer $b$) depends on the analytical 
continuation adopted and it is thus devoid of any intrinsic physical 
value. Its only use is to be integrated to extract the free energy finite 
differences through Eq.~(\ref{intfb}), which are independent of the 
analytical continuation used (for an explicit numerical check see  
\cite{cffmm}) and are physically 
meaningful as far as $b_1$ and $b_2$ are integers. 

Once $\Delta f(B, T)$ has been computed by using Eq.~(\ref{intfb}) we have to 
properly renormalize it, in order to allow for a smooth continuum limit 
extrapolation. The only divergences that do not cancel in the difference 
$\Delta f$ are the $B-$dependent ones and it can be show that such divergences
are temperature independent (see \emph{e.g.} \cite{reg0, reg2}). Motivated by 
this result and by the physical observation that we are interested
in the magnetic properties of the thermal medium and not in those of the 
vacuum, we adopted the renormalization prescription
\begin{equation}\label{subtr}
\Delta f_R (B,T) = \Delta f(B,T) - \Delta f(B,0)\ . 
\end{equation}
From the behavior of $\Delta f_R(B,T)$ for small fields it is possible to 
verify that the medium is linear and, eventually, to extract the value of its 
magnetic susceptibility by using
\begin{equation}\label{intfree2}
\Delta f_R \simeq -\frac{\tilde\chi}{2\mu_0}\mathbf{B}^2 \equiv 
-\frac{\hat\chi}{2}(e\mathbf{B})^2\ .
\end{equation}
$\tilde{\chi}$ is related to the standard SI magnetic susceptibility by the 
relation $\chi=\tilde{\chi}/(1-\tilde{\chi})$ and it is used in order
to properly take into account the fact that in our simulations the medium 
has no back-reaction on the magnetic field (see \cite{cffmm} for
more details). The equivalent of $\tilde{\chi}$ in natural units is 
$\hat{\chi}$ defined by the last equality in Eq.~(\ref{intfree2}) and 
the relation between the two susceptibilities is simply $\hat{\chi}\simeq 10.9 \tilde{\chi}$.

\section{Numerical results}

The method described in the previous section has been applied in \cite{cffmm} to 
the study of the magnetic properties of $N_f=2$ QCD. The theory was discretized 
by using the standard rooted staggered formulation and, although $m_u=m_d$, 
isospin symmetry is explicitly broken by the interaction with the magnetic field, 
since $q_u=2|e|/3$ and $q_d=-|e|/3$. 

To use Eq.~(\ref{intfb}) we need to measure the observable
\begin{equation}\label{M_def}
M \equiv a^4 \frac{\partial f}{\partial b}=\frac{1}{4 L_t L_xL_yL_z} 
\sum_{q = q_u,q_d} \Big\langle \mathrm{tr}\Big\{ 
\frac{\partial D^{(q)}}{\partial b} {D^{(q)}}^{-1} \Big\} \Big\rangle\ , 
\end{equation}
where $D^{(q)}$ is the Dirac matrix of the charge $q$ fermion and $L_{\mu}$ is the 
lattice extent in the $\mu$ direction. $M$ was evaluated by means of a noisy estimator, 
using for each measure $10$ random vectors. Measures have been performed on 
$\mathcal{O}(10^3)$ configurations generated by the usual RHMC algorithm for each value 
of the parameters used, \emph{i.e.} for values of the pion mass in the range $200-480\,\mathrm{MeV}$ 
and for several values of the lattice spacing (for more details see Tab.~1 of \cite{cffmm}). 
As our reference $T=0$ value for the renormalization subtraction we used the 
result obtained on symmetric lattices.

\begin{figure}[t!]
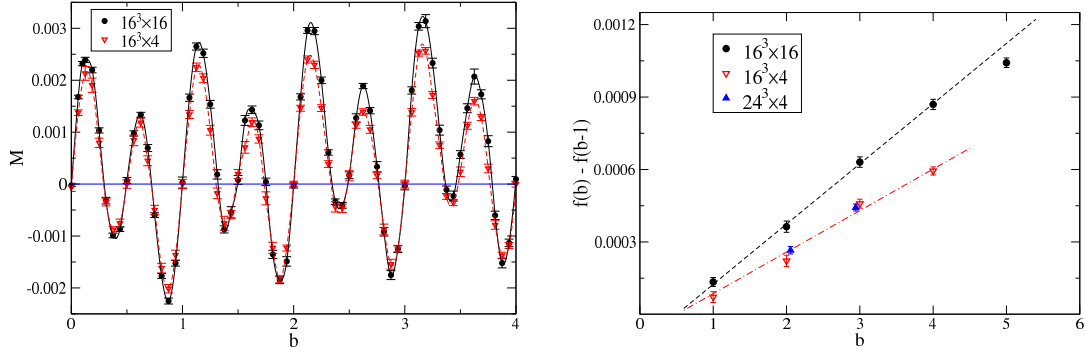

\begin{center}
\includegraphics[width=0.45\columnwidth, clip]{M_m480_a0.188.eps}
\hspace{0.5cm}
\includegraphics[width=0.45\columnwidth, clip]{F_16x16_m480_a0.188.eps}
\end{center}
\caption{Some results obtained with lattice spacing $a\approx 0.188$\,fm and pion 
mass $m_{\pi}\approx 480$\,MeV. (\emph{left}) $M$ computed on $16^4$ and $16^3 
\times 4$ lattices together with third order spline interpolations; (\emph{right}) 
$f(b)-f(b-1)$ computed on $16^4, 16^3\times 4$ and $24^3\times 4$ lattices together 
with the linear fit explained in the text.} \label{MF_fig}
\end{figure}

An example of the results obtained for $M$ is shown in the left panel of 
Fig.~(\ref{MF_fig}). The oscillations in the results are a clear signal of 
the unphysical nature of $M$ and are related to the presence of the unphysical 
string when $b$ is not an integer. Two different harmonics are visible in the result, 
which can be associated to the $u$ and $d$ contributions to $M$. Oscillations are 
nevertheless smooth enough for the result to be numerically integrated. The 
integration, using $16$ determination of $M$ in each quantum, is performed by 
using a spline interpolation and a bootstrap analysis is used to evaluate the 
numerical error. Several tests have been performed by using different integration 
schemes, spline interpolations and number of $M$ determinations; in all the cases 
compatible result are obtained, which shows that the integration procedure is 
very stable (see the Supplementary Material of \cite{cffmm} for more details).

Assuming the relation $a^4\Delta f=c_2 b^2+\mathcal{O}(b^4)$ to hold true for 
integer $b$ values, a convenient way to extract the coefficient $c_2$ is to study 
the differences 
\begin{equation}\label{dDF_fit}
a^4\, (f(b) - f(b-1)) \equiv \int_{b-1}^b M(\tilde b) \mathrm{d}\tilde{b}\, 
\simeq\, c_2\, (2b-1)\ . 
\end{equation}
This is convenient since in this way we do not need to compute 
$\partial f(b)/\partial b$ on the whole $[0,b]$ interval but only on some quanta. 
This strategy also presents the advantage that the integration error does not 
correlate the measures on different quanta, whose estimates are thus statistically 
independent of each other.

Some data for these free energy differences, together with fits according to 
Eq.~(\ref{dDF_fit}), are shown in the right panel of Fig.~(\ref{MF_fig}). From this
figure it can be seen that the fit nicely works for small enough magnetic field, 
and thus the strongly interacting medium is linear, while for greater $b$ values 
deviations from Eq.~(\ref{dDF_fit}) are visible. In all the cases we limited ourself 
to the study of the leading linear term, which is the one needed to extract the 
magnetic susceptibility.

\begin{figure}[t!]
\begin{center}
\includegraphics[width=0.6\columnwidth, clip]{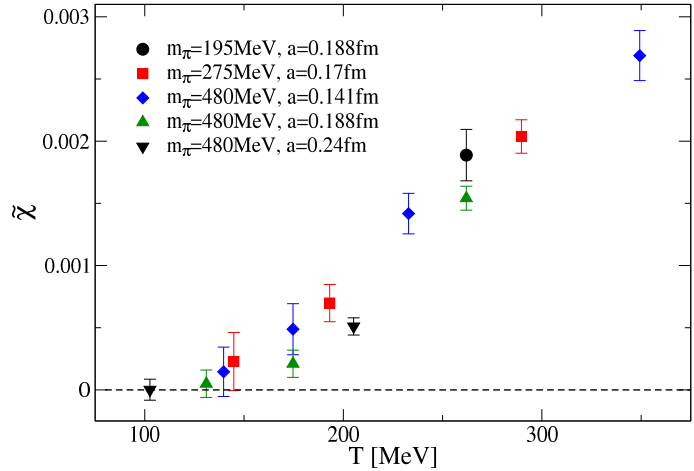}
\end{center}
\caption{Final results for the $\tilde{\chi}$ magnetic susceptibility 
in SI units.}\label{final_fig}
\end{figure}

By following this strategy both for the finite $T$ and the $T=0$ data we arrive to
the relation $a^4\Delta f_R=c_{2\, R} b^2+\mathcal{O}(b^4)$, where 
$c_{2\, R}=c_2(T)-c_2(T=0)$. The last step needed to extract the magnetic susceptibility 
is just a conversion into physical units: 
\begin{equation}
\tilde{\chi} = - \frac{|e|^2 \mu_0 c}{18 \hbar \pi^2}\, (L_xL_y)^2\, c_{2\,R} \qquad
\quad \hat{\chi} = - \frac{1}{18 \pi^2}\, (L_xL_y)^4\, c_{2\,R}\ .
\end{equation}
The data obtained for $\tilde{\chi}$ are shown in Fig.~(\ref{final_fig}) 
(for the numerical data see Tab.~1 of \cite{cffmm}) and, since
$|\tilde{\chi}|\ll 1$, we have $\chi\approx\tilde{\chi}$, where $\chi$ is the usual
magnetic susceptibility in SI units. 

Fig.~(\ref{final_fig}) displays several interesting features. First of all we notice
that data do not show any significant dependence on the lattice spacing and only
slightly depend on the value of the pion mass. The magnetic susceptibility is everywhere 
non negative, so we have shown that strongly interacting matter at finite temperature 
behaves as a paramagnetic medium. Moreover the value of the magnetic susceptibility 
in the explored range is of the same order of magnitude of that of strongly 
paramagnetic ordinary materials, like \emph{e.g} liquid oxygen.
Another interesting feature of Fig.~(\ref{final_fig}) is the strong increasing 
of the magnetic susceptibility in the neighbourhood of the deconfinement crossover, 
which for the masses used in this work is located in the range $160-170\mathrm{MeV}$:
in the low temperature phase, data are much smaller that the ones in the deconfined 
phase, and in fact they are compatible with zero within errors.

\section{Discussion and conclusions}

In this proceeding we presented the method introduced in \cite{cffmm} to study the magnetic 
properties of the finite temperature strongly interacting matter and the first results obtained by
applying it to the case of $N_f=2$ staggered fermions. This method is theoretically well founded and
completely non-perturbative, all the systematic errors can be analyzed independently and 
they turned out to be well under control (see \cite{cffmm} for more details). 
The results obtained by means of other approaches (see \cite{levkova} and \cite{reg_lattice, reg_last}) give a 
qualitatively similar picture of the dependence of the magnetic susceptibility on the temperature.

The most natural extension of the numerical results presented in this proceeding is the use of 
improved discretizations and physical quark masses. This has been done in \cite{cffmm_last} 
by using $2+1$ flavors, a tree-level Symanzik improved action for the gauge fields, a stout smearing 
improvement for the staggered fermions and physical values for the $u$, $d$ and $s$ masses. 
Higher values of the magnetic susceptibility are obtained in this new setting, but the main 
features of Fig.~(\ref{final_fig}) remains unaltered. It was however possible to obtain a better
signal to noise ratio in the confided phase and, in particular, to explicitly display the paramagnetic
behaviour of the low temperature phase. These results are in good quantitative agreement with the 
analogous ones reported in \cite{reg_last}. Of particular phenomenological relevance could be the 
observation that, near deconfinement, the magnetic contribution to the pressure is a relevant fraction 
($\sim 15\%$ for $|e|B\sim 0.1\,\mathrm{GeV}^2$, $\sim 50\%$ for $|e|B\sim 0.2\,\mathrm{GeV}^2$)
of the thermal contribution (see \cite{cffmm_last}) and could possibly induce even-by-event fluctuations 
(see \cite{reg_last}).

\vspace{0.5cm}

\noindent {\bf Acknowledgements:} 
We thank E.~D'Emilio, E.~Fraga and S.~Mukherjee for useful discussions. 
Numerical computations have been performed on computer facilities
provided by INFN, in particular on two GPU farms in Pisa and Genoa 
and on the QUONG GPU cluster in Rome.

\end{document}